\documentclass[twocolumn,floatfix,preprintnumbers,nofootinbib,superscriptaddress]{revtex4}

\usepackage{ulem}
\usepackage{bm}
\usepackage{times}
\usepackage{amssymb,amsbsy,amsmath,amsfonts}
\usepackage{graphicx}
\usepackage{float}
\usepackage{color}
\usepackage{morefloats}
\usepackage{rotating}
\usepackage{srcltx}
\usepackage{slashed}
\usepackage{subfigure}
\usepackage{multirow}
\usepackage{verbatim}
\usepackage{hyperref}
\usepackage{tabularx}
\usepackage{braket}

\usepackage[outdir=./]{epstopdf}

\usepackage{graphicx}
\usepackage{amsmath}
\usepackage{amsfonts}
\usepackage{amssymb}
\usepackage{color}
\usepackage{multirow}
\usepackage{mathrsfs}
\usepackage{gensymb}

\usepackage{booktabs}  
\usepackage{threeparttable}

\usepackage{subfigure}

\newcommand\be{\begin{eqnarray}}
\newcommand\ee{\end{eqnarray}}

\begin{document}

\title{The electromagnetic form factors and spin polarization of $\Lambda_c^+$ in the process $e^+ e^- \to \Lambda^+_c \bar{\Lambda}^-_c$}

\author{Cheng Chen}~\email{chencheng@impcas.ac.cn}
\affiliation{Institute of Modern Physics, Chinese Academy of Sciences, Lanzhou 730000, China}
\affiliation{School of Nuclear Sciences and Technology, University of Chinese Academy of Sciences, Beijing 101408, China}

\author{Bing Yan}~\email{yanbing@impcas.ac.cn}
\affiliation{Institute of Modern Physics, Chinese Academy of Sciences, Lanzhou 730000, China}

\author{Ju-Jun Xie}~\email{xiejujun@impcas.ac.cn}
\affiliation{Institute of Modern Physics, Chinese Academy of Sciences, Lanzhou 730000, China}
\affiliation{School of Nuclear Sciences and Technology, University of Chinese Academy of Sciences, Beijing 101408, China}
\affiliation{Southern Center for Nuclear-Science Theory (SCNT), Institute of Modern Physics, Chinese Academy of Sciences, Huizhou 516000, Guangdong Province, China}

\begin{abstract}

The total cross sections of the process $e^+ e^-\to \Lambda_c^+ \bar{\Lambda}_c^-$ close to the threshold are calculated within the vector meson dominance model. It is found that the theoretical results can describe the current experimental measurements. The nontrivial near-threshold energy dependence of the total cross sections of the process $e^+ e^-\to \Lambda_c^+ \bar{\Lambda}_c^-$ can be well reproduced by considering the contributions from charmonium-like states and the Coulomb factor. In particular, the results for the angular distribution parameters about the differential cross section are consistent with the experiments from BESIII Collaboration. In addition, the relative phase $\Delta \Phi$ of the electromagnetic form factors was given, and the spin polarization of $\Lambda_c^+$ is predicted at center-mass energy $4.7 \ \rm GeV$. It is hopeful to provide a new perspective on the characteristics of the charmed baryon $\Lambda_c^+$.

\end{abstract}

\maketitle
\section{Introduction}

Electromagnetic form factors (EMFFs) are essential to understanding the electromagnetic structure of baryons~\cite{1,2,3}. In the single photon exchange approximation, the EMFFs are as a function of $q^2$, and the physical region is categorized into the space-like region ($q^2<0$) and time-like region ($q^2>0$), according to the transferred four-momentum squared $q^2$ of the exchanged virtual photon. In the space-like region, the EMFFs are measured by the scattering channel $e^-B\to e^-B$ ($B$ stands for a baryon)~\cite{4,5,6}, and reflect the electromagnetic spatial distribution of baryons. However, due to the difficulties in producing stable and high-quality hyperon beams, it is challenging to study the EMFFs of hyperons in the spacelike region. In the time-like region, the EMFFs are investigated in the annihilation process $e^-e^+ \to B\bar{B}$ ($\bar{B}$ stands for an antibaryon), which is the main process to study the EMFFs of unstable baryons~\cite{7,8,9,10,11,12,13,14}. Moreover, the EMFFs in the time-like region were considered as the time evolution of the charge and magnetic distributions inside baryon~\cite{15}. On the other hand, the spin polarization and asymmetry parameters of $\Lambda$, $\Sigma$, and $\Xi$ hyperons are also measured in the $e^+e^- \to \Lambda \bar{\Lambda}$~\cite{16,17}, $e^+ e^- \to \Sigma \bar{\Sigma}$~\cite{14,18}, and $e^+ e^- \to \Xi \bar{\Xi}$~\cite{19,20,21} reactions, respectively. The spin polarization of hyperons are related to the relative phase between their electric $G_E$ and magnetic $G_M$ form factors.

The $\Lambda_c^+$ is the lightest baryon with a single-charm quark, and its quantum numbers spin-parity are $J^P = 1/2^+$. It has only weak decay modes. In 2008, the cross section of the reaction $e^+e^-\to \Lambda_c^+\bar{\Lambda}_c^-$ was first measured by Belle Collaboration~\cite{22}. The corresponding cross section demonstrates a nontrivial energy dependence in the vicinity of the reaction threshold, which shows a resonance structure around 4.63 GeV. In 2018, BESIII Collaboration reported more precise data near the reaction threshold and found the threshold enhancement effect in the process $e^+e^-\to \Lambda_c^+\bar{\Lambda}_c^-$~\cite{23}. However, the BESIII measurement of the $e^+ e^- \to \Lambda^+_c \bar{\Lambda}^-_c$ reaction from the threshold to $4.6$ GeV implies a different energy-dependence trend of the total cross sections~\cite{23}. Recently, high-precision data for the reaction was measured by BESIII Collaboration~\cite{24}, where the flat cross sections around 4.63 GeV are obtained and no indication of the resonant structure about $Y(4630)$, as reported by the Belle Collaboration~\cite{22}. In addition to the total cross section of $e^+e^-\to \Lambda_c^+\bar{\Lambda}_c^-$ reaction, the magnetic form factor $|G_M|$, the ratio $|G_E/G_M|$, and the angular distribution parameter $\alpha $ of $\Lambda^+_c$ are also measured in Ref.~\cite{24}. In particular, it was found that the energy dependence of the ratio $|G_E/G_M|$ for the $\Lambda^+_c$ baryon reveals an oscillation feature.

Before the new measurements of the $e^+ e^- \to \Lambda^+_c \bar{\Lambda}^-_c$ reaction by the BESIII Collaboration~\cite{24}, the final state interactions of the $\Lambda^+_c \bar{\Lambda}^-_c$ pair and the possible $X(4660)$ state were considered in the theoretical studies on the reaction of $e^+e^-\to \Lambda_c^+\bar{\Lambda}_c^-$~\cite{25,26,27}, and the near-threshold enhancement observed by the Belle Collaboration was reproduced. It was pointed that a virtual state lying below but very close to the $\Lambda^+_c \bar{\Lambda}^-_c$ mass threshold, located at $4.566 \pm 0.007$ GeV, causes the threshold enhancement~\cite{26}. Moreover, based on the Belle data~\cite{22} and the BESIII data~\cite{23}, both the time-like and space-like EMFFs of the $\Lambda_c^+$ were investigated by the vector meson dominance model (VMD) in Ref.~\cite{28}, in which six charmonium states below the $\Lambda^+_c \bar{\Lambda}^-_c$ mass threshold were included in the model. The new results of Ref.~\cite{24} indicate that there are more complicated reaction dynamics behind in the process of $e^+ e^- \to \Lambda^+_c \bar{\Lambda}^-_c$. Therefore, further studies on this reaction are needed. 

With the new measurements by the BESIII Collaboration~\cite{24}, in Ref.~\cite{29}, it is pointed out that the contact potential in the final state $\Lambda_c^+\bar{\Lambda}_c^-$ rescattering plays an important role in the enhancement near the threshold. Up to $4.75$ GeV, the available data on the total cross sections of $e^+ e^- \to \Lambda^+_c \bar{\Lambda}^-_c$, modulus of the individual EMFFs of $\Lambda^+_c$, and their ratio can be well described~\cite{29}. Besides, in Ref.~\cite{30}, with the VMD model as in Ref.~\cite{28}, the $e^+ e^- \to \Lambda^+_c \bar{\Lambda}^-_c$ cross sections and the time-like EMFFs of $\Lambda_c^+$ were investigated by including four charmonium-like states that have been experimentally discovered in recent years~\cite{31,32}, called $\psi(4500)$, $\psi(4660)$, $\psi(4790)$, and $\psi(4900)$. It was found that these charmonium-like states are crucial to the nonmonotonic structures of the $\Lambda_c^+$ EMFFs and enhancement near the reaction threshold~\cite{30}. Especially, the so-called oscillation behavior of the ratio $|G_E/G_M|$ for the $\Lambda^+_c$ baryon discovered by the BESIII Collaboration~\cite{24} can be also well reproduced. This gives a natural production mechanism for the $e^+ e^- \to \Lambda^+_c \bar{\Lambda}^-_c$ reaction.

Here, in this work, motivated by the planed new measurements~\cite{33,34} for the $e^+ e^- \to \Lambda^+_c \bar{\Lambda}^-_c$ reaction at threshold,~\footnote{This is because the $\Lambda^+_c$ can be reconstructed via the Cabibbo-favored weak decay of $\Lambda^+_c \to p K^- \pi^+$, even it was produced at rest.} we will focus on the details of electromagnetic form factors of $\Lambda_c^+$ based on our previous study~\cite{30}, including the total cross section near the threshold and modulus of electric form factor $G_E$. Furthermore, we will also study the angular distribution parameter $\alpha$, the relative phase $\Delta \Phi$ between the electric form factor $G_E$ and magnetic form factor $G_M$, and the spin polarization of $\Lambda_c^+$ in the reaction of $e^+e^-\to \Lambda_c^+\bar{\Lambda}_c^-$.  

The paper is organized as follows. In next section, we give the definition of electromagnetic form factors in the reaction of electron-positron annihilation into a pair of baryon and antibaryon, and from this, we show the associated spin polarization of the produced charmed baryon $\Lambda^+_c$ in the $e^+ e^- \to \Lambda^+_c \bar{\Lambda}^-_c$ reaction in Sec. III. Then, we show the theoretical results and discussions in Sec. IV. Finally, a short summary is given in the last section.

\section{Electromagnetic Form Factors}

Generally, the scattering amplitude of the reaction $e^+e^-\to \gamma^* \to B\Bar{B}$ ($B$ stands for a baryon with spin-parity $J^P = 1/2^+$) within the one photon exchange approximation can be written as
\begin{eqnarray}
    \mathcal{A} = \frac{e^2}{q^2}\Bar{v}_e(k_1)\gamma^\mu u_e(k_2)\braket{B(p_1)\bar{B}(p_2)|J^{em}_\mu|0},
\end{eqnarray}
where $q = k_1 + k_2 = p_1 + p_2 $, and the matrix element $\braket{B(p_1)\bar{B}(p_2)|J^{em}_\mu|0}$, for the electromagnetic current conservation and Lorentz invariance, could be parameterized by two independent form factors,
\begin{align}
    &\braket{B(p_1)\bar{B}(p_2)|J^{em}_\mu|0} \nonumber \\
    &= \bar{u}_B(p_1)\left[F_1(q^2)\gamma_\mu + i\frac{F_2(q^2)}{2M}\sigma_{\mu\nu}{q^\nu}  \right]v_{\bar{B}}(p_2),
\end{align}
where $M$ is the mass of the baryon $B$, and $F_1(q^2)$ and $F_2(q^2)$ are the so-called Dirac and Pauli form factors, respectively, with the normalizations $F_1(0) = Q_B$, $F_2(0) = \kappa_B$. Here, $Q_B$ and $\kappa_B$ denote the charge and anomalous magnetic moment of the baryon $B$. In this way, the electromagnetic information of the baryon $B$ is absorbed into the two scalar functions $F_1(q^2)$ and $F_2(q^2)$.

Based on the above formula, the differential cross section of the reaction in the center of mass frame is
\begin{align}
    \frac{d\sigma}{d\Omega} = & \frac{\alpha^2_{em}\beta C}{4s} \times \nonumber \\
    & \left[|G_M|^2(1+\cos^2 \theta) + \frac{4M^2}{s}|G_E|^2 \sin^2 \theta \right], \label{eq:dsigdomega}
\end{align}
where $s=q^2$ is the invariant mass squared of the $e^+ e^-$ system, and $\theta$ is the scattering angle between the momentum of the initial electron and the final baryon $B$. $\alpha_{em} = e^2/4\pi$ is the electromagnetic fine structure constant, and $\beta = \sqrt{1-4M^2/s}$ is the velocity of baryon $B$. In general, the factor $C$ is the $S$-wave Sommerfeld–Gamow factor corresponding to the final state Coulomb interaction of charged baryons~\cite{35}, which will lead to an unvanished cross section at the threshold, $C = y/(1-e^{-y}),\  y = 2\pi\alpha_{em}M/(\beta\sqrt{s})$. $C\equiv 1$ for neutral baryons. Considering the factor $C$, it is expected that the cross section of $e^+ e^- \to B\bar{B}$ reaction is nonzero at the reaction threshold for charged baryon-antibaryon pair of $B\bar{B}$. In addition, $G_E$ and $G_M$ are the electric and magnetic form factors, respectively, which is the linear combination of $F_1$ and $F_2$,
\begin{align}
    G_E(q^2) &= F_1(q^2) + \tau F_2(q^2), \label{GE} \\
    G_M(q^2) &= F_1(q^2) +  F_2(q^2), \label{GM}
\end{align}
where $\tau = q^2/4M^2$. At reaction threshold, $\tau = 1$, thus $G_E = G_M = F_1 + F_2$.

On the other hand, the differential cross section, shown in Eq.~\eqref{eq:dsigdomega}, can be also expressed in a more compact form
\begin{align}
    \frac{d\sigma}{d\Omega} = N(1+\alpha \cos^2\theta) ,  \label{df}
\end{align}
with 
\begin{align}
    N &= \frac{\alpha^2_{em}\beta C}{4s}\left(|G_M|^2 + \frac{4M^2}{s}|G_E|^2\right), \\ 
    \alpha &= \frac{s|G_M|^2 - 4M^2|G_E|^2}{s|G_M|^2 + 4M^2|G_E|^2}, \label{alpha}
\end{align}
with $\alpha$ depending on the ratio of the absolute values of the electric and magnetic form factors and it is satisfying $-1 \leq \alpha \leq 1$. According to Eq.~(\ref{df}), one can see that the angular distribution of the differential cross section is only dependent on the parameter $\alpha$. On the theoretical side, once we get the electromagnetic form factors $G_E$ and $G_M$, we can easily obtain the parameter $\alpha$ from Eq.~\eqref{alpha}.

Within the above ingredients, and after integrating over the solid angle $\Omega$, the total cross section of the $e^+ e^- \to \Lambda^+_c \bar{\Lambda}^-_c$ reaction is obtained as
\begin{align}
    \sigma  = \frac{4\pi \alpha^2_{em} \beta C}{3s} \left( |G_M(q^2)|^2 + \frac{2M^2_{\Lambda^+_c}}{s}|G_E(q^2)|^2 \right).\label{cross}
\end{align}
Note that, on the experimental side, by combining the angular distributions and the total cross sections, the modulus of $G_E$ and $G_M$ for $\Lambda^+_c$ could be extracted.~\footnote{Here after, $G_E$ and $G_M$ stand for the electric and magnetic form factors of $\Lambda^+_c$.} 

Due to the nonperturbative nature of quantum chromodynamics (QCD) theory in the energy regime of hadrons, an exact theoretical description of their internal structure has not been achieved within the framework of QCD. On the theoretical side, within the extend vector meson dominance model, the Dirac and Pauli form factors $F_1(s)$ and $F_2(s)$ can be phenomenologically parametrized as (more details can be found in Ref.~\cite{30}),
\begin{eqnarray}
    F_1(s) &=& \frac{1}{(1-\gamma s)^2} \left( f_1 +  \sum_{i=1}^4 \beta_i B_{R_i}  \right ),  \label{F1} \\
    F_2(s) &=& \frac{1}{(1-\gamma s)^2} \left( f_2 B_{R_1}  + \sum_{i=2}^4 \alpha_i B_{R_i} \right), \label{F2}
\end{eqnarray}
with parameter $\gamma = 0.147\pm 0.017 \ \rm GeV^{-2}$ and 
\begin{eqnarray}
B_{R_i} = \frac{M_{R_i}^2}{M_{R_i}^2 - s - i M_{R_i} \Gamma_{R_i}}, 
\end{eqnarray}
where we take $R_1 \equiv \psi(4500)$ with mass $M_{R_1} = 4500 \ \rm MeV$,\; $R_2 \equiv \psi(4660)$ with mass $M_{R_2} = 4670 \ \rm MeV$ and width $\Gamma_{R_2} = 115\ \rm MeV$~\cite{36}; $R_3 \equiv \psi(4790)$ with $M_{R_3} = 4790 \ \rm MeV $ and $ \Gamma_{R_3} = 100\ \rm MeV$~\cite{32}; $R_4 \equiv \psi(4900)$ with $M_{R_4} = 4900 \ \rm MeV $ and $ \Gamma_{R_4} = 100\ \rm MeV$~\cite{37,38,39}. Note that since the mass of $\psi(4500)$ is below the $\Lambda^+_c \bar{\Lambda}^-_c$ threshold, for its width we take the Flatt\'e type~\cite{40} 
\begin{eqnarray}
    \Gamma_{\psi(4500)} = \Gamma_0 + g_{\Lambda_c}\sqrt{\frac{s}{4} - M_{\Lambda_c^+}^2},
\end{eqnarray}
with $\Gamma_{0} = 125\ \rm MeV$~\cite{31} and the coupling $g_{\Lambda_c} = 1.173\pm 0.259 $ as determined in Ref.~\cite{30}. 

At $s = 0$ and setting the widths $\Gamma_{R_i} = 0$, with the constraints $G_E = 1$ and $G_M = \mu_{\Lambda^+_c} =  0.24\ \hat{\mu}_N$, the coefficients $f_1$ and $f_2$ in Eq.~\eqref{F1} and \eqref{F2} are obtained as 
\begin{align}
    f_1 &= 1 - \beta_1 - \beta_2 - \beta_3 - \beta_4 ,\\
    f_2 &= \mu_{\Lambda_c^+} - 1 - \alpha_2 - \alpha_3 - \alpha_4 ,
\end{align}
with $\beta_1 = 1.883\pm 0.484$, $\beta_2 = -1.101\pm0.302$, $\beta_3 = -0.439\pm0.194$, $\beta_4 = -0.141\pm 0.097$ and $\alpha_2 = 1.089\pm 0.297$, $\alpha_3 = 0.438\pm 0.192$, $\alpha_4 = 0.133 \pm 0.096$ the model parameters, which are fitted to the experimental data on the total cross sections, the ratio $|G_E/G_M|$ and $G_M$ about the $e^+e^- \to \Lambda^+_c \bar{\Lambda}^-_c$ reaction.

\section{Spin polarization of the produced $\Lambda^+_c$ in the process of $e^+ e^- \to \Lambda^+_c \bar{\Lambda}^-_c$}

In the time-like region, $G_E$ and $G_M$ are complex. The relative phase $\Delta \Phi$ between $G_E$ and $G_M$ is an observable physical quantity, 
\begin{align}
G_E/G_M  = e^{i \Delta \Phi} |G_E/G_M|.
\end{align}
Actually, $\Delta \Phi$ can be extracted experimentally from the spin polarization of the final $\Lambda^+_c$ in the reaction $e^+e^- \to \Lambda^+_c \Bar{\Lambda}^-_c$. However, it is experimentally difficult to measure the spin polarization of baryons directly. The unexpected observation of spin polarization in $e^+ e^- \to J/\psi \to \Lambda \bar{\Lambda}$ by the BESIII Collaboration~\cite{41} has opened us new perspectives for such studies~\cite{18,42,43,44,45,46}.

\begin{figure*}[htbp]  
    \centering
    \includegraphics[scale=1.]{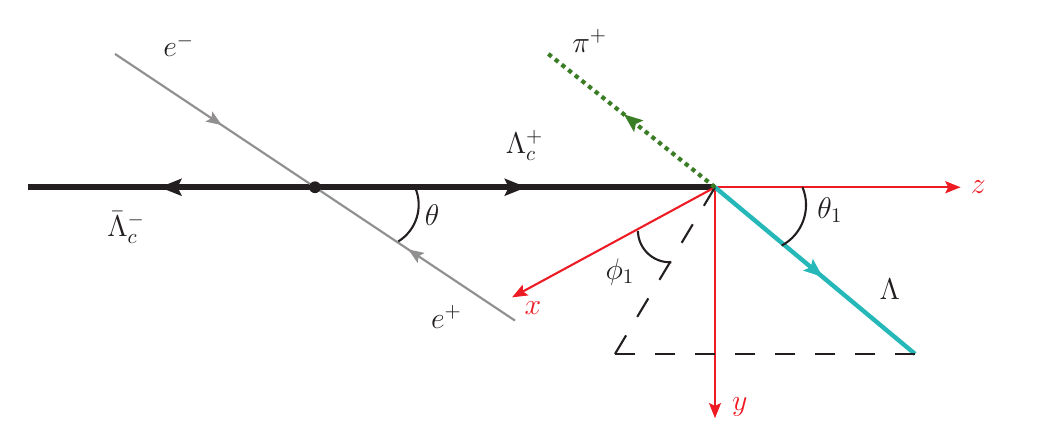}
    \caption{The definition of the coordinate system in the reaction  $e^+e^-\to \Lambda^+_c (\to \Lambda + \pi^+)\bar{\Lambda}^-_c (\to \bar{\Lambda} + \pi^-)$, where the decay of $\bar{\Lambda}_c^-$ to $\bar{\Lambda}\pi^-$ is not shown for simply. The solid angle $\Omega_\Lambda = (\theta_1,\phi_1)$ of $\Lambda$ is defined in the center of mass frame of $\Lambda_c^+$, and $\Omega_{\bar{\Lambda}} = (\theta_2,\phi_2)$ in $\bar{\Lambda}^-_c$ with the same coordinate system.}   \label{fig:axis}
\end{figure*}

Usually, the spin direction of $\Lambda^+_c$ is reconstructed by the final particle angular distribution of the $\Lambda^+_c$  weak decays, for example, $\Lambda^+_c \to \Lambda + \pi^+$. Taking into account the final $\Lambda^+_c$ and $\bar{\Lambda}^-_c$ weak decays, the whole process as $e^+e^-\to \Lambda^+_c (\to \Lambda + \pi^+)\bar{\Lambda}^-_c (\to \bar{\Lambda} + \pi^-)$, its differential distribution can be expressed as~\cite{47,48}
\begin{align}
& \mathcal{W}(\xi) = 1 + \alpha \mathcal{T}_5(\xi) + \nonumber \\ 
 & \alpha_{\Lambda_c^+}\alpha_{\bar{\Lambda}_c^-}\left[\mathcal{T}_1(\xi) + \sqrt{1-\alpha^2}\cos(\Delta\Phi)\mathcal{T}_2(\xi) + \alpha \mathcal{T}_6(\xi)\right] \nonumber \\ 
 &+ \sqrt{1-\alpha^2}\sin (\Delta\Phi)[\alpha_{\Lambda_c^+} \mathcal{T}_3(\xi) + \alpha_{\bar{\Lambda}_c^-} \mathcal{T}_4(\xi) ], \label{wxi}
\end{align}
where $\xi = (\theta, \Omega_\Lambda, \Omega_{\bar{\Lambda}})$, and $\Omega_\Lambda, \Omega_{\bar{\Lambda}},$ is the momentum direction of the baryon $\Lambda$ and $\bar{\Lambda}$ in the rest frame of $\Lambda^+_c$ and $\bar{\Lambda}^-_c$, respectively. And here, $\alpha_{\Lambda_c^+}$ is the decay asymmetry parameter for the  $\Lambda^+_c \to \Lambda + \pi^+$ decay, and $\alpha_{\bar{\Lambda}_c^-}$ for the $\bar{\Lambda}_c^- \to \bar{\Lambda} + \pi^- $.~\footnote{If $CP$ is conserved in the charge conjugate decay, it will lead to $\alpha_{\Lambda_c^+} = -\alpha_{\bar{\Lambda}_c^-}$. Here, we take $CP$ is conserved. The distribution function $\mathcal{W}(\xi) $ can be used to extract separately $\Lambda^+_c$ and $\bar{\Lambda}^-_c$ decay asymmetry parameters, and thus allowing to a direct test of $CP$ violation in the hyperon weak decays.} In addition, these angular functions $\mathcal{T}_i ~(i=1,2, \dots, 6)$ are given as following
\begin{align}
    \mathcal{T}_1(\xi)&=\sin^2 \theta \sin \theta_1 \sin \theta_2 \cos \phi_1 \cos \phi_2 + \cos^2 \theta \cos \theta_1 \cos \theta_2, \nonumber \\
    \mathcal{T}_2(\xi)&=\sin \theta \cos \theta (\cos \theta_2 \sin \theta_1  \cos \phi_1 + \cos \theta_1 \sin \theta_2 \cos \phi_2),  \nonumber \\
    \mathcal{T}_3(\xi)&=\sin \theta \cos \theta \sin \theta_1 \sin \phi_1, \nonumber \\
    \mathcal{T}_4(\xi)&=\sin \theta \cos \theta \sin \theta_2 \sin \phi_2, \nonumber \\
    \mathcal{T}_5(\xi)&=\cos^2 \theta,  \nonumber \\
    \mathcal{T}_6(\xi)&=\cos \theta_1 \cos \theta_2 - \sin^2 \theta \sin \theta_1 \sin \theta_2 \sin \phi_1 \sin \phi_2, \nonumber   
\end{align}
where $(\theta_1$,$\phi_1) = \Omega_\Lambda$, $(\theta_2$,$\phi_2) = \Omega_{\bar{\Lambda}}$. Note that the coordinate system is defined as: the $z$ axis direction is along the momentum $\vec{p}$ of the emitted baryon $\Lambda^+_c$, and $y$ axis is perpendicular to the scattering plane along the direction $\vec{p}\times \vec{k}$, with $\vec{k}$ being the electron momentum, and $x$ axis is determined by the right-hand coordinate system, as presented in Fig.~\ref{fig:axis}. 

The terms $1+ \alpha \mathcal{T}_5(\xi)$ in Eq.~(\ref{wxi}) is spin-independent, corresponding to the differential cross section angular distribution in Eq.~(\ref{df}). The second term multiplied by $\alpha_{\Lambda_c^+} \alpha_{\bar{\Lambda}_c^-}$ represents the spin entanglement of $\Lambda^+_c$ and $\bar{\Lambda}^-_c$. While the last term stands for the spin polarization of the charmed baryon $\Lambda^+_c$ along the $y$ axis. High statistic data samples close to threshold for the $e^+ e^- \to \Lambda^+_c \bar{\Lambda}^-_c$ reaction enabled studies of the angular distributions of $\Lambda^+_c$ in the $e^+ e^-$ center-of-mass system.

Experimentally, the information about spin polarization of charmed baryon $\Lambda^+_c$ in the joint reaction $e^+e^-\to \Lambda^+_c (\to \Lambda + \pi^+)\bar{\Lambda}^-_c (\to \bar{\Lambda} + \pi^-)$ at a certain energy $\sqrt{s}$ is mainly obtained by measuring the following angular distribution~\cite{16,41,49},
\begin{align}
 M(\cos \theta) & = \frac{\alpha_{\Lambda_c^+} - \alpha_{\bar{\Lambda}_c^-}}{3+\alpha} \sqrt{1-\alpha^2} \nonumber \\
 & \times \sin(\Delta \Phi)\sin\theta \cos \theta . \label{M}
\end{align}
One can see that the $M(\cos \theta)$ will be vanished at $\alpha = \pm 1$ or the relative phase $\Delta\Phi = 0$, otherwise, it reaches its maximum value at $\theta =  \pi/4$ and vanishes at $\theta = 0$, $\pi/2$, and $\pi$. In general, in the time-like region, $G_E$ and $G_M$ are complex, thus $\Delta\Phi$ is not zero and $\alpha$ cannot reach $\pm 1$, which indicates that the produced $\Lambda^+_c$ is polarized, even the initial states $e^+$ and $e^-$ are both unpolarized. And the spin polarization of $\Lambda^+_c$ is directed along the normal to the $e^- e^+ \to \Lambda^+_c \bar{\Lambda}^-_c$ scattering plane which is defined as $\vec{p} \times \vec{k}$. It is worthy to mention that if the $y$ axis direction is defined by $\vec{k}\times \vec{p}$, there will be an additional minus in the angular distribution function Eq.~(\ref{wxi}), i.e. replacing $\sin \theta$ with $-\sin \theta$.

\section{Numerical results and discussions}

\begin{figure}[htbp]  
	\centering
	\includegraphics[scale=0.40]{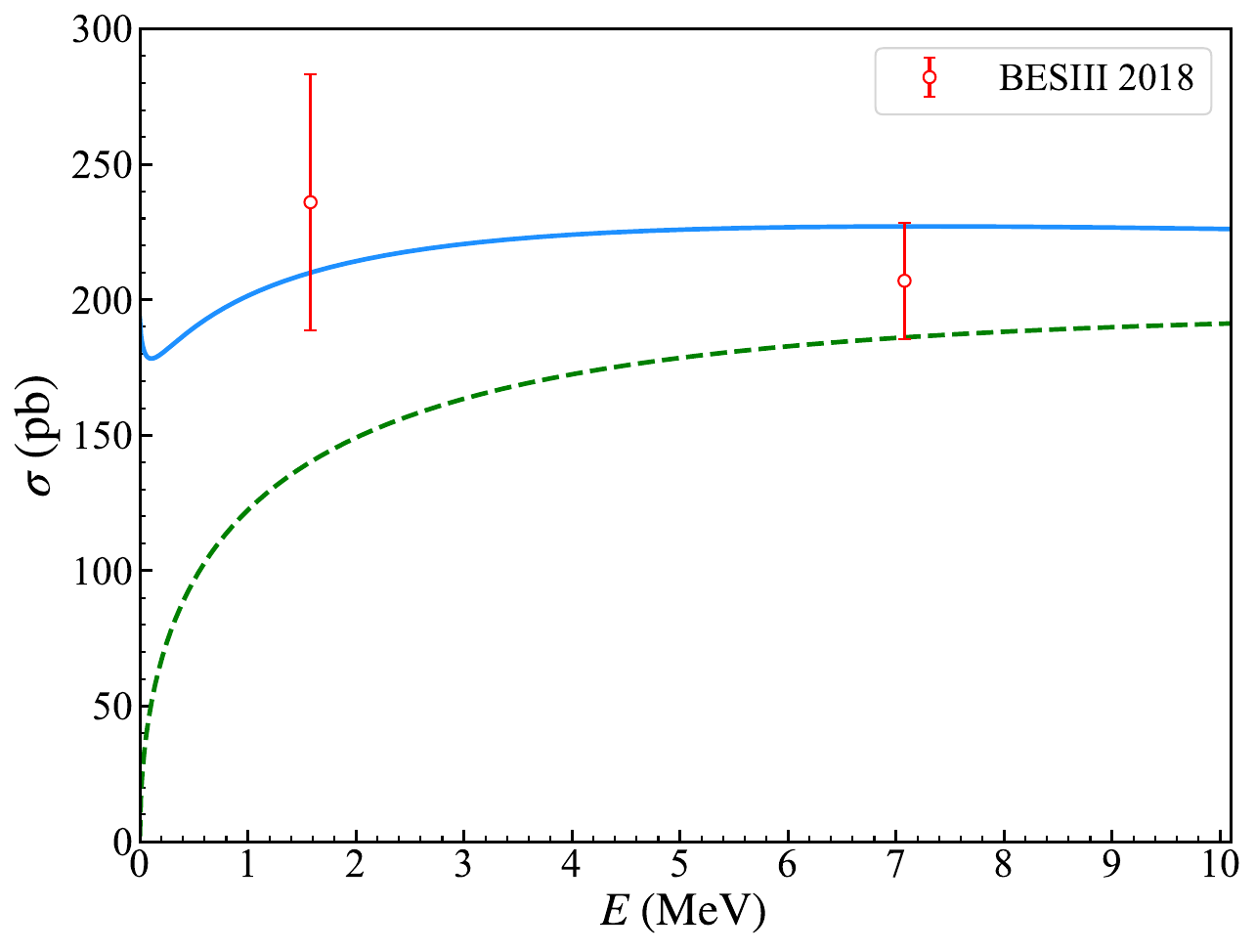}
	\caption{The obtained total cross sections of $e^+ e^- \to \Lambda^+ \bar{\Lambda}^-_c$ reaction as a function of the excess energy $E$ comparison with experimental data taken from BESIII Collaboration~\cite{23}. The blue solid line represents the theoretical results that include the Sommerfeld–Gamow factor $C$, and the green dashed line excludes the factor $C$.}
	\label{fig:CS}
\end{figure}

The Driac and Fauli form factors $F_1(q^2)$ and $F_2(q^2)$ form factors of the $\Lambda_c^+$ were constructed within the VMD model, in which four charmonium-like states were included, called $\psi(4500)$, $\psi(4660)$, $\psi(4790)$, and $\psi(4900)$. Then one can easily get the electromagnetic form factors of $G_E(q^2)$ and $G_M(q^2)$, and also the experimental observables. The model parameters are fixed by fitting them to the experimental data. By taking into account these vector excited states, it is successfully explained the flat behavior of the total cross section from the threshold to $\sqrt{s} = 4.67\ \rm GeV$ for the process $e^+e^-\to \Lambda_c^+\bar{\Lambda}_c^-$~\cite{30}. Here, we present the detailed structure of the cross section near the reaction threshold, as shown in Fig.~\ref{fig:CS} by the blue solid line, where the quantity $E$ is defined as $E \equiv \sqrt{s} - 2M_{\Lambda_c^+}$. One can see that the theoretical results are in agreement with the experimental data taken from BESIII Collaboration~\cite{23}, and the flat behavior of the total cross sections can be well reproduced. Moreover, it is found that the obtained value of the total cross section at the reaction threshold ($E=0$) is about 193 $\rm nb$, and $G_E(E=0) = G_M(E=0) = 0.92- 0.71i $.

To see how important of the factor $C$, we plot also the numerical results without including the factor $C$ in Fig.~\ref{fig:CS} by the green dashed line. One can see that the factor $C$ affects only at the energy region very close to the reaction threshold~\cite{26}. Moreover, its affect decreases very quickly as the reaction energy growing. For energies above a few MeV of the reaction threshold, the impact of this Sommerfeld–Gamow factor is not very important. However, it does provide a nonzero cross section at the reaction threshold.

Because the total cross sections of $e^+e^- \to \Lambda^+_c \bar{\Lambda}^-_c$ reaction can be measured at the reaction threshold, it is expected that more experimental measurements can be done to test our model calculations shown in Fig.~\ref{fig:CS}. In fact, the BESIII detector has collected scan data at fourteen c.m. energies from threshold to 4.95 GeV with relatively large statistics~\cite{33,45}. Thus, it is possible that a high precision production cross-section line-shape of the $e^+ e^- \to \Lambda^+_c \bar{\Lambda}^-_c$ reaction will be produced soon. These future experimental data will serve as crucial inputs for theoretical calculations concerning different production mechanism of $\Lambda^+_c$.

\begin{figure}[htbp]
	\centering
	\includegraphics[scale=0.40]{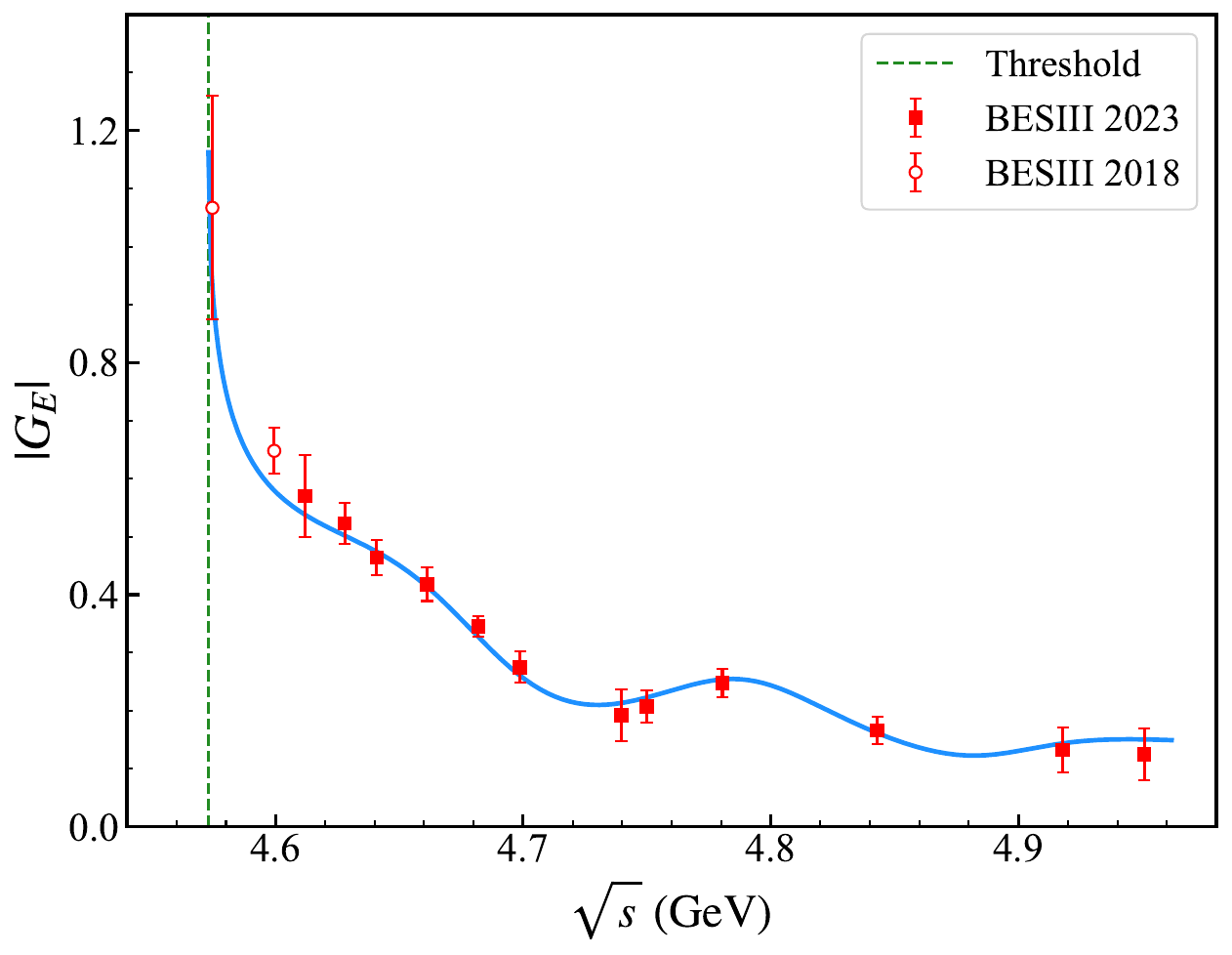}
	\caption{The obtained electric form factor $|G_E|$ compared with experimental data taken from Refs.~\cite{23,24}. The green vertical dashed curves represent the $e^+ e^- \to \Lambda^+_c \bar{\Lambda}^-_c$ kinematic reaction threshold.}
	\label{fig:GE}
\end{figure}

Next, in Fig.~\ref{fig:GE}, the obtained modulus of the electric form factor $G_E(q^2)$ as a function of $\sqrt{s}$ are shown. One can see that the model can well describe the experimental data. In particular, the date point very close to the kinematic threshold can be also reproduced. On the other hand, there is a sizable bump structure around $\sqrt{s} = 4.78\ \rm GeV$ in the line shape of $|G_E(q^2)|$ which is attributed to the charmonium-like state $\psi(4790)$~\cite{32}. Further studies on this state both on experimental and theoretical sides are mostly welcome~\cite{50}.

\begin{figure}[htbp]
	\centering
	\includegraphics[scale=0.40]{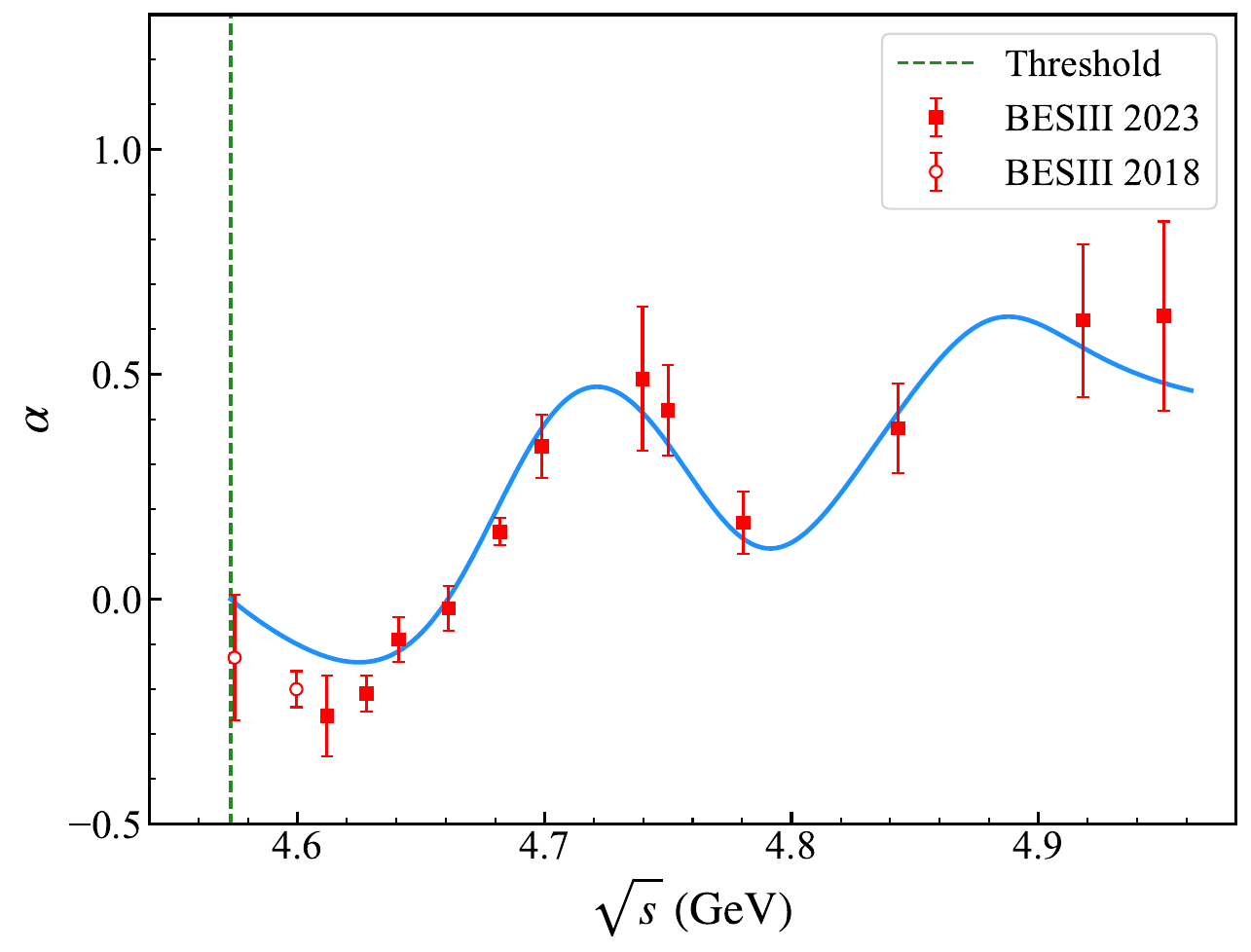}
	\caption{As in Fig.~\ref{fig:GE}, but for the theoretical results about the angular distribution parameter $\alpha$ compared with the experimental data from BESIII Collaboration~\cite{23,24}.}
	\label{fig:Alpha}
\end{figure}

The theoretical results for the angular distribution parameter $\alpha$ are presented in Fig.~\ref{fig:Alpha}, and compared with the experimental data taken from BESIII Collaboration~\cite{23,24}. One can see that our theoretical results are in agreement with the experimental data especially in the c.m. energies from $4.64$ to $4.90\ \rm GeV$. Still, close to the reaction threshold the theoretical results do not match the experimental values so well. However, more precise data around the reaction threshold are also needed. On the other hand, to improve the theoretical calculations near the reaction threshold, it might need to introduce new $\Lambda_c^+\bar{\Lambda}_c^-$ bound states~\cite{51}.

As shown in Fig.~\ref{fig:Alpha}, the line shape of $\alpha$ shows an oscillatory increasing behavior, which is a natural result according to Eq.~(\ref{alpha}), since the ratio of $|G_E/G_M|$ has a damped oscillation~\cite{30}. This so-called oscillating behavior is because of the production of the vector excited states. In addition to these studies on the EMFFs for the light baryons in Refs.~\cite{52,53,54,55,56,57,58,59,60,61,62,63,64,65}, it is concluded that the non-monotonic structures observed in the line shape of the $e^+e^- \to B\bar{B}$ total cross sections can be naturally explained by including the contributions from the excite vectors within the vector meson dominance model.

\begin{figure*}[htbp]  
	\centering
	\includegraphics[scale=0.40]{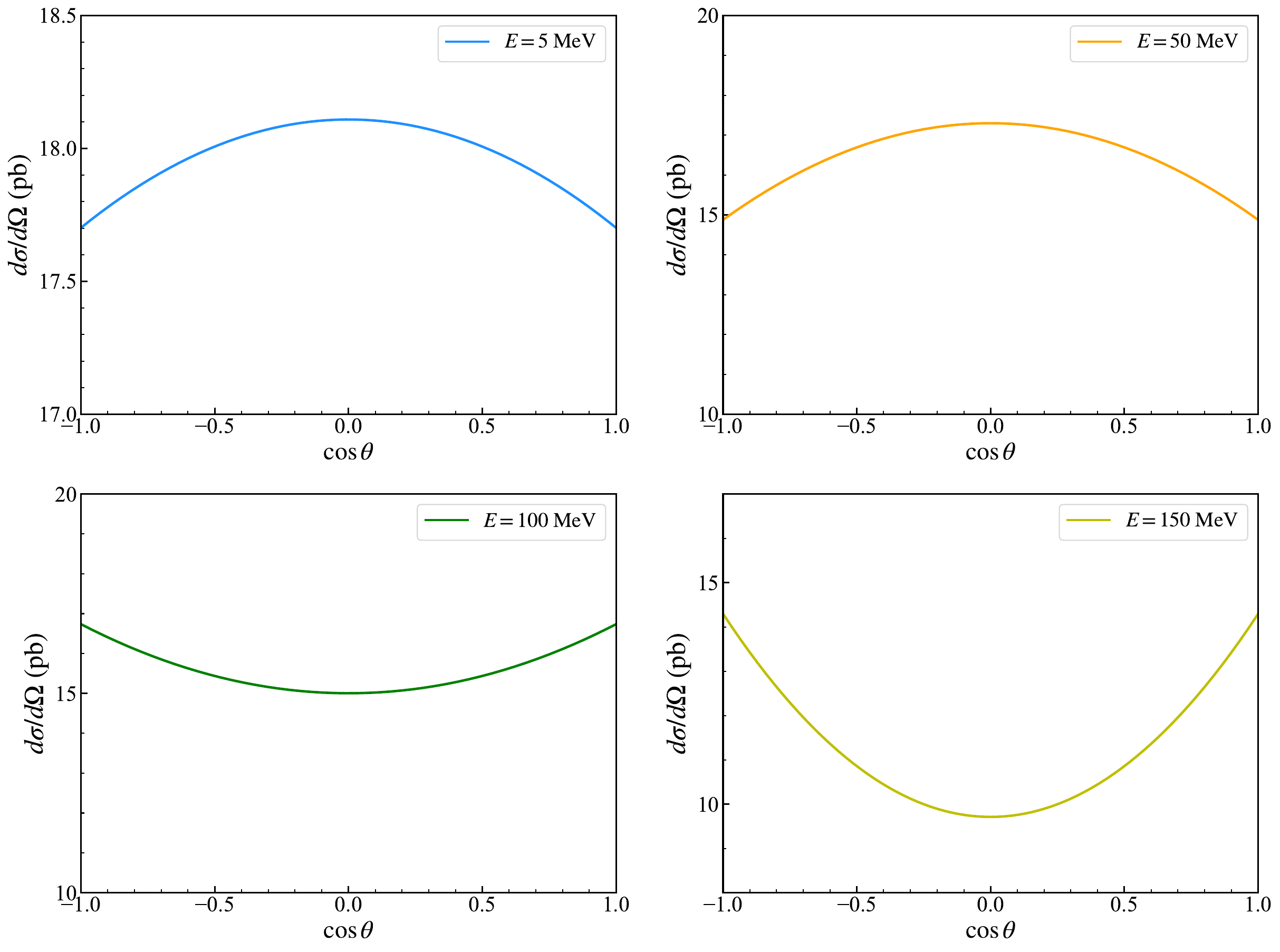}
	\caption{The obtained differential cross sections of the process $e^+ e^- \to \Lambda^+ \bar{\Lambda}^-_c$ at excess energy $E =5$, $50$, $100$, and $150$ MeV.}
	\label{fig:DCS}
\end{figure*}

Besides, with the obtained electromagnetic form factors or the parameter $\alpha$, the differential cross sections of $e^+e^- \to \Lambda^+_c \bar{\Lambda}^-_c$ reaction at $E=5$, $50$, $100$, and $150$ MeV can be easily calculated and the theoretical results are shown in Fig.~\ref{fig:DCS}. On the experimental side, the parameter $\alpha$ and the relative phase angle $\Delta\Phi$ between the electric and magnetic form factors of $\Lambda^+_c$ can be extracted from the differential cross sections of $e^+ e^- \to \Lambda^+ \bar{\Lambda}^-_c$ reaction. Again, more experimental measurements about the differential cross sections from the experimental side are needed to test the theoretical calculations here and determine the individual electromagnetic form factors of $\Lambda^+_c$.

\begin{figure}[htbp]
    \centering
    \includegraphics[scale=0.40]{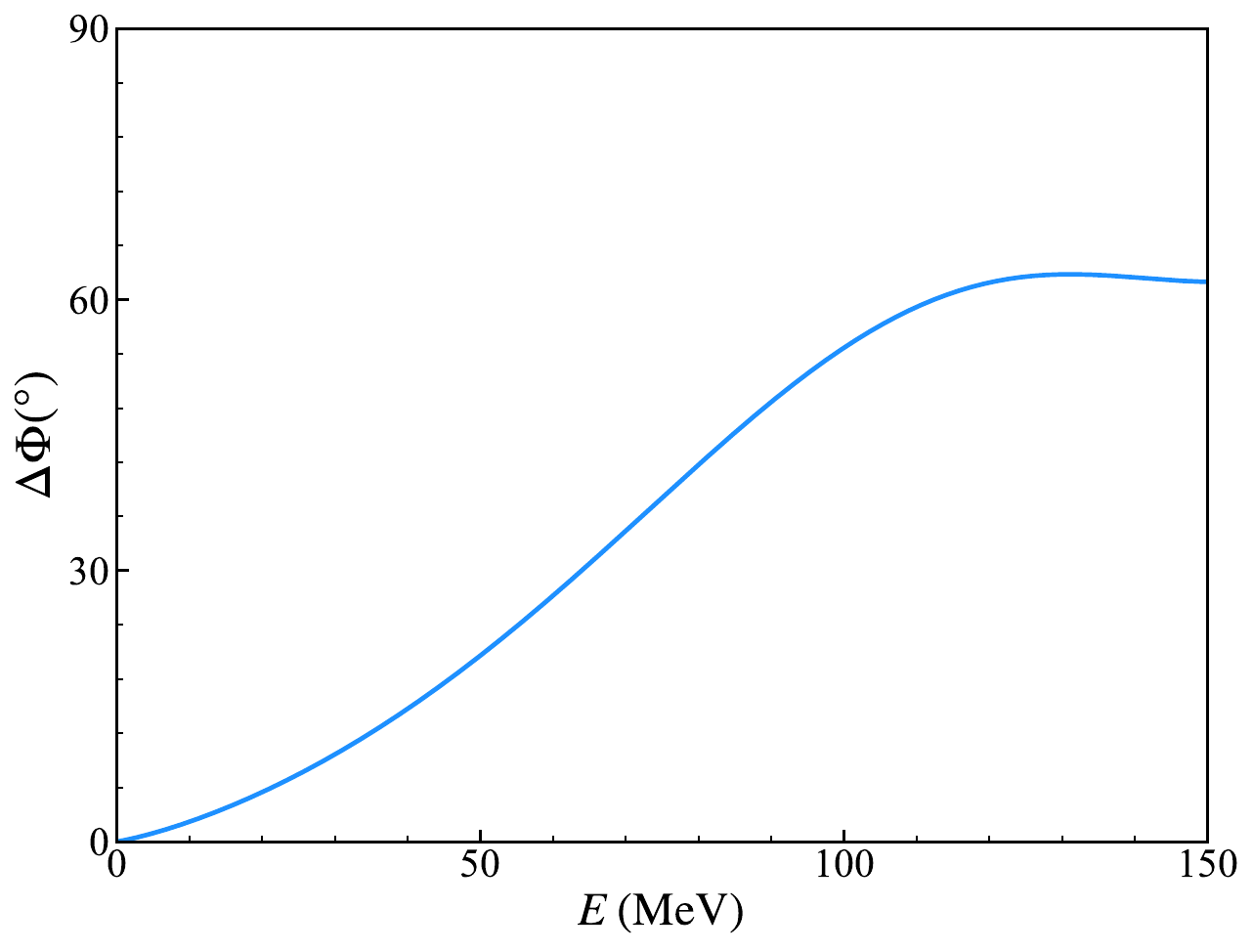}
    \caption{The obtained relative phase $\Delta \Phi$ between $G_E$ and $G_M$ as a function of $E$.}
    \label{fig:Angle}
\end{figure}

The obtained relative phase $\Delta \Phi$ between electric $G_E$ and magnetic $G_M$ form factors are shown in Fig.~\ref{fig:Angle}. At the reaction threshold $E = 0$, the relative phase $\Delta \Phi$ equals zero because $G_E$ is exactly equal to $G_M$ according to Eqs.~(\ref{GE}) and (\ref{GM}). As the energy $E$ increases, the relative phase increases and remains constant at about $60\degree$ around $E = 120 \ \rm MeV$. It is worth mentioning that it is difficult to get the relative phase $\Delta \Phi$ because the experimental cross sections depend on the modulus of $G_E$ and $G_M$. To get this phase, one needs to measure the angular distributions of the $e^+e^- \to \Lambda^+_c \bar{\Lambda}^-_c$ reaction. In fact, the relative phase $\Delta \Phi$ could be measured from the spin angular distribution of the baryon, which usually was reconstructed from their weak decays. Meanwhile, the baryon spin polarization is self-analyzed in their weak decays.

\begin{figure}[htbp]
    \centering
    \includegraphics[scale=0.40]{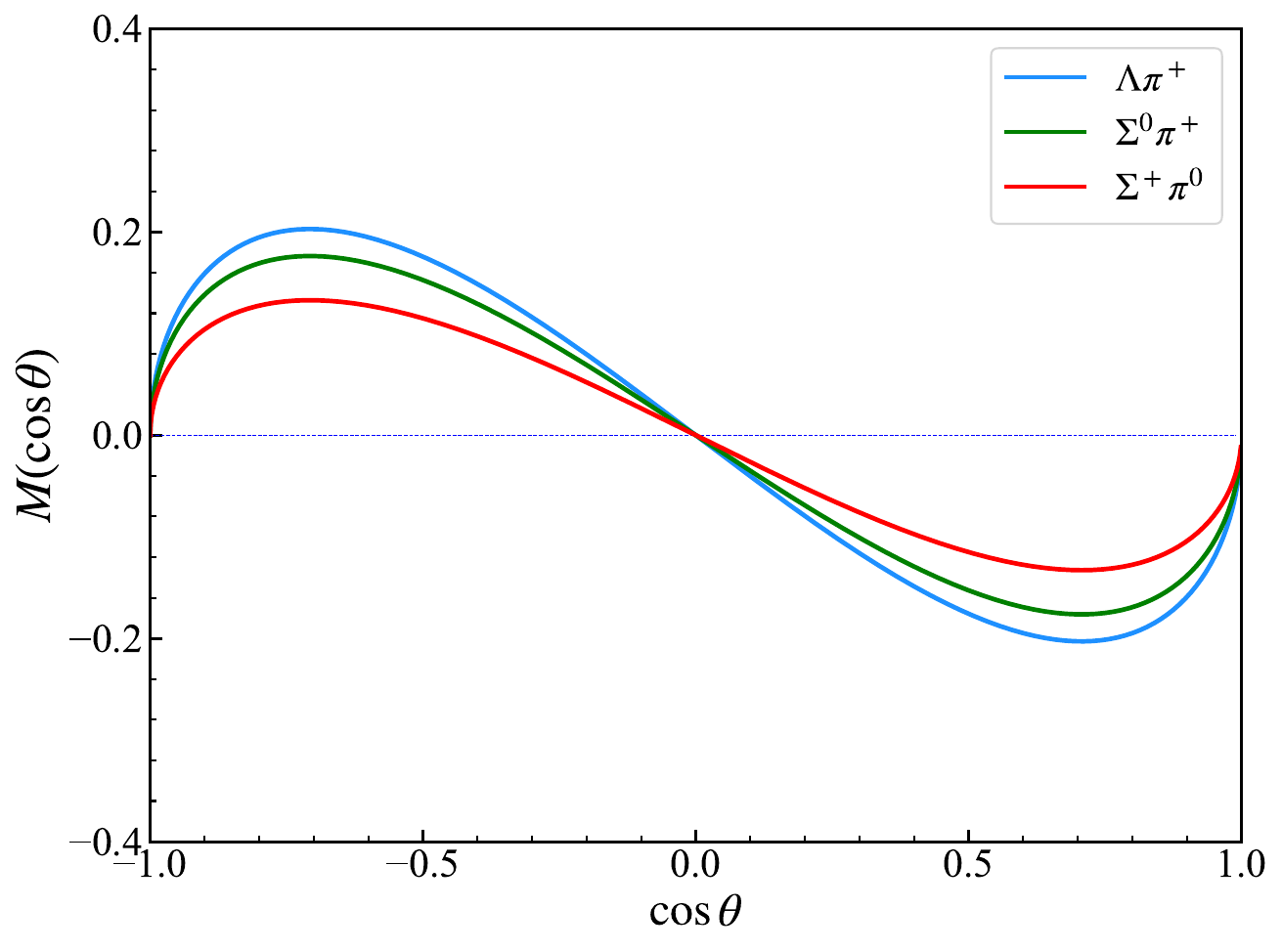}
    \caption{The obtained moments $M(\cos \theta)$ at $ \sqrt{s} = 4.7\ \rm GeV$. The blue line stands for the result of the decay of the final state $\Lambda_c^+$ to $\Lambda\pi^+$, the green line for $\Lambda_c^+ \to \Sigma^0\pi^+$ and the red line for $\Lambda_c^+ \to \Sigma^+\pi^0$, while the decay of $\bar{\Lambda}_c^-$ is the charge conjugation of $\Lambda_c^+$, respectively.}
    \label{fig:Mcos}
\end{figure}

Following Ref.~\cite{47} and according to the obtained form factors $F_1$ and $F_2$ within the VMD model, we can get the scattering amplitude of $e^+e^- \to \Lambda^+_c \bar{\Lambda}^-_c$ reaction, from where we can study the spin polarization of $\Lambda^+_c$. With Eq.~(\ref{M}) where the details about the spin polarization of $\Lambda^+_c$ along the $y$ axis are given for the case of $\Lambda_c^+ \to \Lambda \pi^+$, the distributions of $M(\cos \theta)$ at $ \sqrt{s} = 4.7\ \rm GeV$ (i.e. $E = 125\ \rm MeV$) are calculated. Because, near this energy point, it possesses a considerably larger relative phase, to such an extent that a more remarkable polarization can be experimentally observed. Furthermore, we have also calculated $M(\cos \theta)$ for the cases of $\Lambda_c^+\to\Sigma^0\pi^+$ or $\Lambda_c^+\to\Sigma^+\pi^0$. These three decay modes have larger decay branching ratios compared to other channels~\cite{66}, and the decay asymmetry parameter $\alpha_{\Lambda^+_c}$ in the three final states $\Lambda\pi^+$, $\Sigma^0\pi^+$ and $\Sigma^+\pi^0$ are $-0.84\pm0.09$, $-0.73\pm0.18$, $-0.55\pm0.11$, respectively. In Fig.~\ref{fig:Mcos}, we show our numerical results about the $M(\cos \theta)$ in the three decay channels at c.m. energies $\sqrt{s} = 4.7\ \rm GeV$. The blue line stands for $\Lambda_c^+\to\Lambda\pi^+$, the green line for the $\Lambda_c^+\to\Sigma^0\pi^+$ and the red line for the $\Lambda_c^+\to\Sigma^+\pi^0$. It is expected that the theoretical results for the $M(\cos \theta)$ can be measured by the BESIII Collaboration and Belle Collaboration in future experiments or at the planned super tau charm facility~\cite{67}. Anyway, it is worth anticipating that more experimental information will be available in the near future to greatly improve the knowledge about the production mechanism of the charmed baryon $\Lambda^+_c$ in the process of $e^+ e^- \to \Lambda^+_c \bar{\Lambda}^-_c$.

\section{Summary}

In summary, we investigated the electromagnetic form factors of the lightest charmed baryon $\Lambda_c^+$ and the total cross sections of $e^+e^-\to \Lambda_c^+\bar{\Lambda}_c^-$ near threshold based on the extended vector meson dominance model. Using parameterizations of electromagnetic form factors $G_E$ and $G_M$ available, the angular distribution parameter $\alpha$ is also calculated, which is consistent with the experimental data from BESIII Collaboration. Moreover, the modulus of electric form factor $G_E$ and the relative phase $\Delta \Phi$ between $G_E$ and $G_M$ were presented. The relative phase is associated with the spin polarization of the produced baryon in the $e^+ e^- \to \Lambda^+_c \bar{\Lambda}^-_c$ reaction. With the obtained relative phase and the parameter $\alpha$, the moments $M(\cos \theta)$ are obtained in three $\Lambda_c^+$ decay channels, $\Lambda_c^+ \to \Lambda\pi^+$, $\Lambda_c^+\to \Sigma^0\pi^+$ and $\Lambda_c^+ \to \Sigma^+\pi^0$. Since the channel $\Lambda_c^+ \to \Lambda\pi^+$ has a larger decay asymmetry parameter compared to the other two decay modes, it is more likely that significant phenomena will be observed in the process $e^+e^-\to \Lambda_c^+(\to \Lambda\pi^+)\bar{ \Lambda}_c^-(\to \bar{\Lambda}\pi^-)$. It is expected that these theoretical results obtained here can be tested by future experiments, such as BESIII and Belle II.

Finally, we would like to stress that thanks to the important role played by the excited charmonium states in the $e^+ e^- \to \Lambda^+_c \bar{\Lambda}^-_c$ reaction, more precise experimental measurements for this reaction, such as Belle II, BESIII, and STCF, etc., can be used to improve our knowledge on the properties of some charmonium states, which are at present poorly known. More experimental information will be available in the near future, which will greatly improve our knowledge about the production mechanism of the process $e^+ e^- \to \Lambda^+_c \bar{\Lambda}^-_c$.

\section*{Acknowledgements}

We would like to thank Prof. Pei-Rong Li and Prof. Xiong-Fei Wang for useful discussions. This work is partly supported by the National Key R\&D Program of China under Grant No. 2023YFA1606703, and by the National Natural Science Foundation of China under Grant Nos. 12075288, 12435007 and 12361141819. It is also supported by the Youth Innovation Promotion Association CAS.

\end{document}